\documentclass[twocolumn,showpacs,preprintnumbers,amsmath,amssymb]{revtex4}

\usepackage{graphicx}
\usepackage{dcolumn}
\usepackage{bm}

\begin{document}


\title{Overcritical state in superconducting round wires sheathed by iron}

\author{Alexey V. Pan}
\email{pan@uow.edu.au}
\author{Shixue Dou}
\affiliation{Institute for Superconducting and Electronic Materials, University of Wollongong, \\ Northfields Avenue, Wollongong, NSW 2522, Australia}

\date{April 20, 2003}

\begin{abstract}
Magnetic measurements carried out on MgB$_2$ superconducting round wires have shown that the critical current density $J_c(B_a)$ in wires sheathed by iron can be significantly higher than that in the same bare (unsheathed) wires over a wide applied magnetic field $B_a$ range. The magnetic behavior is, however, strongly dependent on the magnetic history of the sheathed wires, as well as on the wire orientation with respect to the direction of the applied field. The behavior observed can be explained by magnetic interaction between the soft magnetic sheath and superconducting core, which can result in a redistribution of super-currents in the flux filled superconductor. A phenomenological model explaining the observed behavior is proposed.
\end{abstract}
\pacs{74.25.Ha, 74.25.Sv, 84.71.Mn}
 
\maketitle

\section{Introduction}

The current-carrying ability of superconductors has been long debated since practical applications became viable. The dissipation-free current flow in type-II superconductor can exist in either the Meissner state within the magnetic field penetration depth $\lambda$ or in the Shubnikov (mixed) state as long as the magnetic vortices are pinned. The field operational range of the Meissner state is usually too small and super-currents too weak to satisfy practical needs. Hence an enormous effort has been made to find ways to reinforce the vortex pinning, which determines the maximum dissipation-free currents in the superconductors in the Shubnikov state. An additional method for further enhancing the current-carrying ability of superconducting strips was suggested in a series of recent theoretical works by Genenko {\it et al.} \cite{gen1,gen}. It was shown that thin superconductors placed in a soft magnetic environment with a high magnetic permeability can exhibit transport currents higher than those which would be normally obtained in the critical state. This situation was referred to as the {\em overcritical} state. A maximum overcritical current $I^{oc}$ higher than the critical current $I_c$ by a factor of $\sim 7$ was estimated to be likely obtained in some practical cases \cite{gen}. However, the strips require a peculiar geometrical shape of the magnetic environment within which the strips should be situated \cite{gen1,gen}. This obstacle made it difficult to achieve and experimentally test this magnetic method.

New promising horizons for testing and employing the magnetic environment method for enhancement of the current-carrying ability have been opened up by MgB$_2$ superconducting wires sheathed by magnetic materials such as iron (Fe) \cite{joseph,shield,sumpt}. Indeed, round MgB$_2$ superconducting wires sheathed in iron seem to exemplify objects which are quite simple to fabricate and deal with. Previous works have already reported on the observation and benefits of the magnetic shielding effect for round MgB$_2$ wires \cite{joseph,shield}. The shielding effect has been long-known for a shell of highly permeable material \cite{dj}. However, the interaction between the ferromagnetic iron sheath and the superconductor produces numerous prominent new electro-magnetic effects if the wire is placed in an externally applied magnetic field $B_a$ \cite{joseph,shield}. Among the most intriguing are: (i) a sharp drop in $I_c(B_a)$ at small fields and a plateau in the intermediate field region obtained in transport experiments, in which the applied magnetic field was superimposed on the applied transport current \cite{joseph}; (ii) a strong magnetic-history dependence of the magnetization behavior as a function of the applied field observed in magnetic experiments, in which only an external field was applied \cite{shield}.

In this work, we focus on issues related to the magnetic history-dependent effects induced by the interaction of the superconductor and the magnetic sheath in the case where only an external field is applied. In the next section we describe the experimental procedures employed in this work. In Sect.~\ref{er} we provide results obtained for the case of the field applied perpendicular (Sect.~\ref{sub1}) and parallel (Sect.~\ref{sub2}) to the round wire axis. In Section~\ref{discoc}, which is the main part of the paper, we discuss the magnetic history-dependent results and propose a model to explain the observation of an overcritical state. The behavior of virgin zero-field cooled magnetization curves is considered in Sect.~\ref{discv}. The influence of thermo-magnetic flux-jump instabilities is described in Sect.~\ref{discfj}. In the last section we summarize the results obtained in this work.

\section{Experimental details}

The magnetization measurements on MgB$_2$ superconducting wires sheathed by iron were performed with the help of a Quantum Design MPMS SQUID magnetometer within an applied magnetic field range of $|B_a| \le 5$~T at different temperatures and two field orientations with respect to the cylindrical wire axis: perpendicular and parallel. The results of the measurements on sheathed wires are compared to those carried out under absolutely the same conditions but on bare wires. A bare wire was obtained by careful and thorough mechanical removal of the iron sheath from the wire so that the superconducting volume of the core remained nearly the same.

The Fe-sheathed MgB$_2$ superconducting wires were made by employing the powder-in-tube technique and using an {\it in-situ} Mg+2B powder reaction approach for forming the MgB$_2$ superconducting core \cite{pan}. The wires investigated in this work were similar to those described in Ref.~\onlinecite{shield}.
 
\section{Experimental results}\label{er}

\begin{figure}
\includegraphics[scale=0.31]{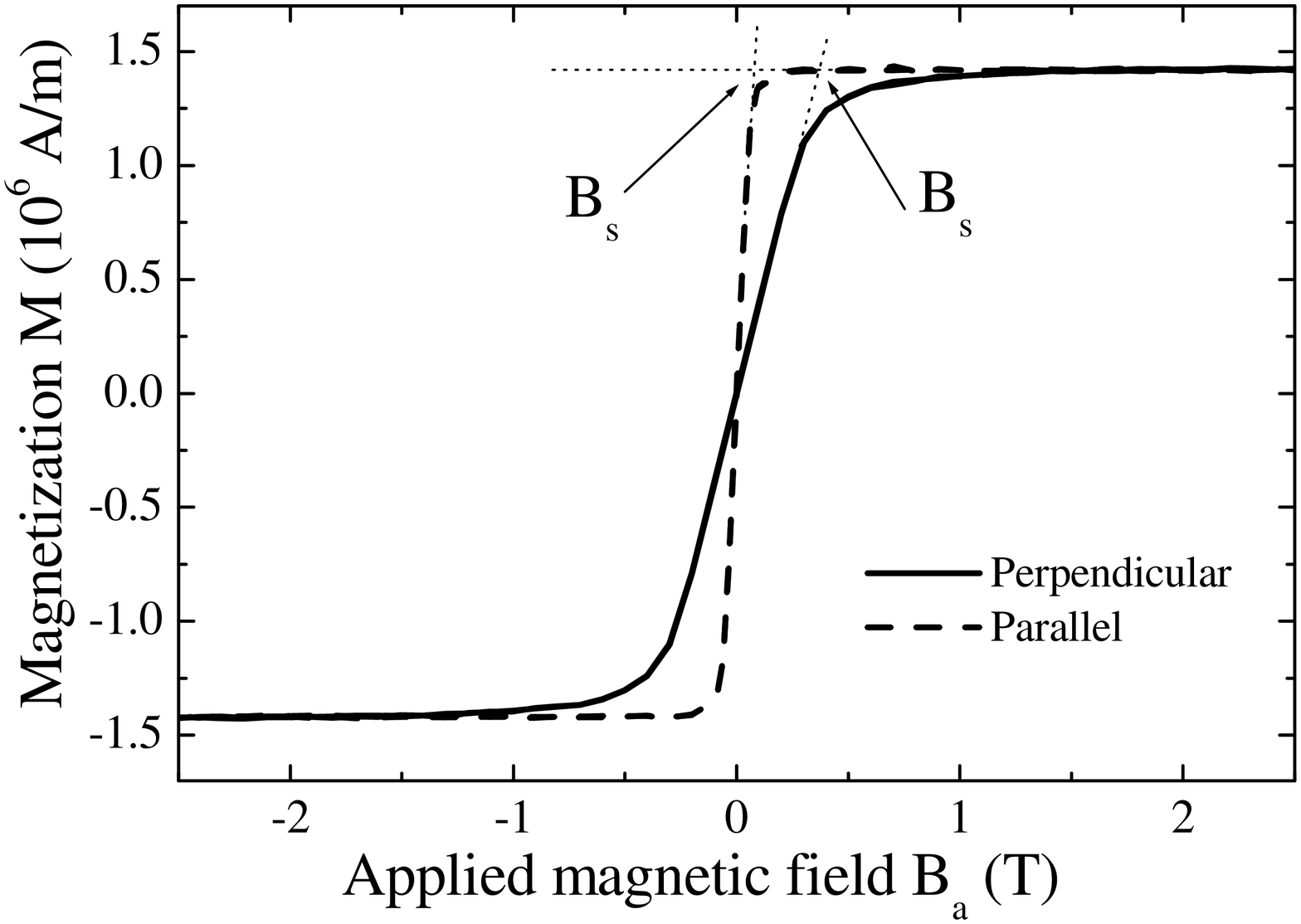}
\caption{\label{iron}The magnetization hysteresis loops of the iron sheath measured above $T_c$ at $T = 43$~K for perpendicular and parallel field orientations. The saturation fields $B_s$ are defined at the crossing of the extensions (dotted lines) to the linear parts of the curves.}
\end{figure}

The magnetic response of the Fe-sheathed wires is dominated by the ferromagnetic signal of iron not only above, but also below the critical temperature $T_c$ \cite{shield}. $T_c \simeq 38.6~K$ has been measured for our samples at $B_a = 2.5$~mT.  To elucidate the superconducting contribution of the core below $T_c$, we have subtracted the Fe-sheath contribution measured above $T_c$ (Fig.~\ref{iron}). The ferromagnetic loop of the sheath contribution exhibits insignificant hysteresis (not visible on the scale in Fig.~\ref{iron}), which is characteristic behavior for soft magnetic materials. The measured loops for the perpendicular (parallel) field orientation provide the Coercitive field of $|B_{cf}| \simeq 1.1 (1.5)$~mT. The remnant magnetization is $|M_r| \simeq 5 (37) \times 10^3$~A/m. The maximal magnetic permeability of the iron before the saturation is $\mu/\mu_0 \simeq 6 (46)$, where $\mu_0$ is the permeability of free space.

In this section we present the {\em superconducting} contribution of the magnetization curves as a function of the applied magnetic field for the Fe-sheathed MgB$_2$ wire. We compare these results with corresponding curves obtained for the {\it bare} superconducting core having no Fe-sheath, i.e., no Fe-induced signal subtraction was needed. Zero-field cooled (ZFC) and field-cooled (FC) types of magnetization measurements were carried out. The ZFC state of the sample was achieved by cooling it through its critical temperature at $B_a = 0$~T. By FC state we imply that measurement started at $|B_a| = 5$~T, which is either $>> B_{c1}$ at $T < 25$~K or $> B_{\rm irr}$ at $T \ge 25$~K. $B_{\rm irr}$ is defined at the field where the descending and ascending branches of the magnetization merge.

The measured curves are shown in Fig.~\ref{prp} for perpendicular and in Fig.~\ref{prl} for parallel orientations. Every graph for each temperature contains curves measured under the following conditions. (i) ZFC magnetization curves measured from $B_a = 0$~T to 1.0~T for the Fe-sheathed MgB$_2$ wire are plotted by large solid circles. (ii) FC magnetization loops, measured starting from $B_a = 5$~T, sweeping to $B_a = -5$~T and then back to $B_a = 5$~T, are plotted by large open circles. In order to directly visualize the influence of the iron sheath, we carried out the same measurements for the bare wire obtained after the removal of the sheath from the same measured wire. Note that the direct comparison is valid since the superconducting volume of the sample remained nearly unchanged. These results are plotted by notably {\it smaller} (diamond) symbols and thicker connecting lines: (iii) the solid diamond symbols are ZFC measurements performed at the same conditions as in case (i); (iv) the open diamonds are FC measurements obtained as described for case (ii). For clarity, only the positive applied field half of the curves in the first and fourth quadrants are shown, because the other half is inversely symmetrical with respect to the ${\bf m}$-axis and does not contain additional information.

Numerous thermo-magnetic flux-jumps are observed at $T < 15$~K as can be seen in Figs.~\ref{prp} and \ref{prl}. Nevertheless, they do not obscure the analysis of the data not only because the remaining temperature range ($15\,{\rm K} \le T \le T_c$) is sufficiently broad, but also because the overcritical effect discussed in this paper does not significantly change below $T \simeq 30$~K as will become clear from Sect.~\ref{discoc}. The flux-jumps are further discussed in Sect.~\ref{discfj}.

\subsection{Perpendicular orientation}\label{sub1}

\begin{figure*}
\vspace{-0.5cm}
\includegraphics[scale=0.65]{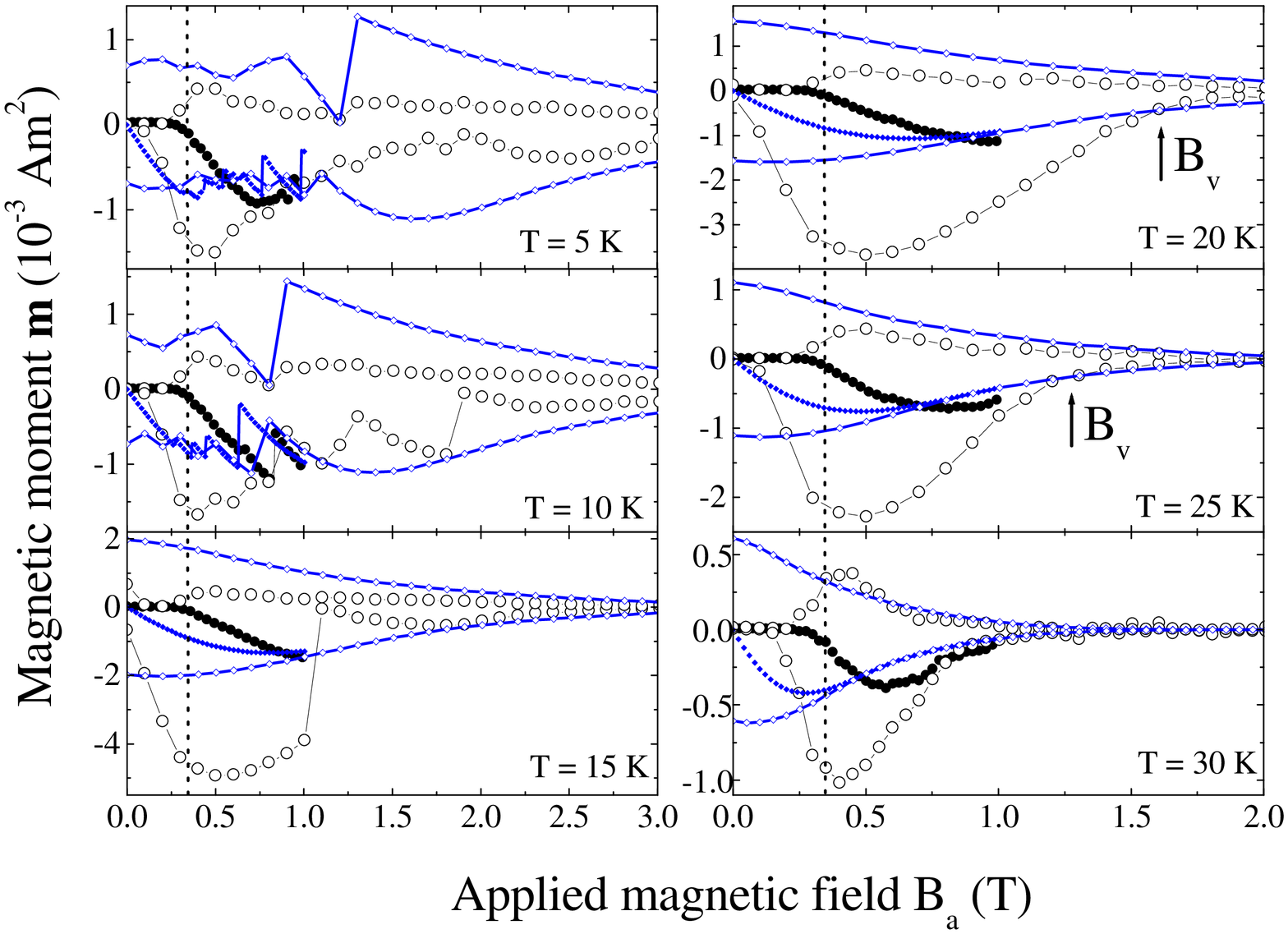}
\caption{\label{prp}The magnetization curves measured for a perpendicular field orientation with respect to the wire axis. The small symbols and thick connecting lines denote curves measured for the bare wire, while large circles and thin connecting lines show results for the Fe-sheathed wire. Open symbols denote FC half-loops, while solid symbols denote ZFC measurements. The dotted lines mark the magnetization saturation field $B_s$ of the iron sheath (see text). The arrows define $B_{\rm v}$.}
\end{figure*}

\begin{figure*}
\vspace{-0.5cm}
\includegraphics[scale=0.65]{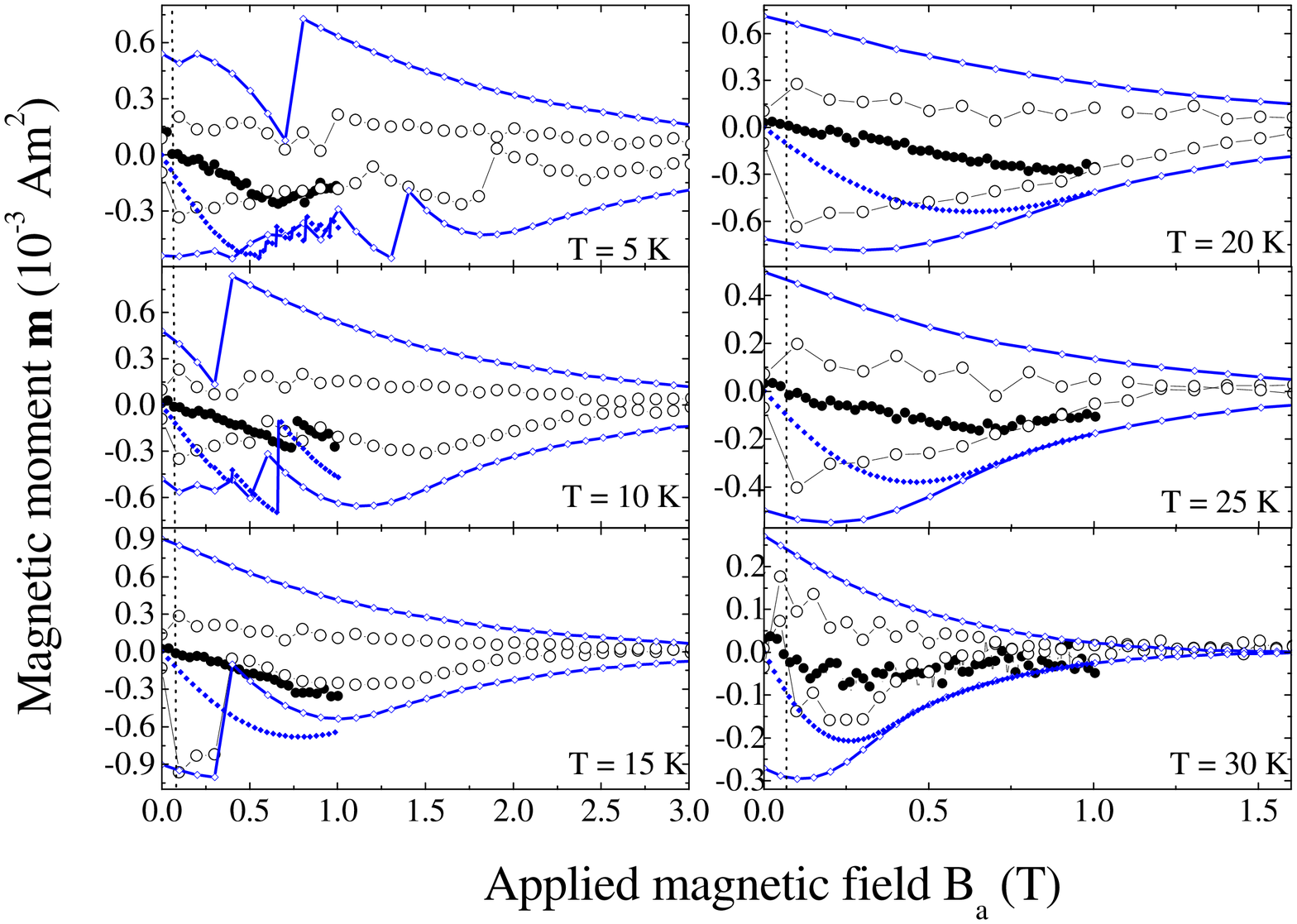}
\caption{\label{prl}The magnetization curves measured for parallel field orientation with respect to the wire axis. The legend is the same as described in the caption for Fig.~\ref{prp}.}
\end{figure*}

The curves shown in Fig.~\ref{prp} exhibit magnetization behavior measured in magnetic fields applied perpendicular to the cylindrical axis of the wire at different temperatures. Strikingly, one finds numerous discrepancies between the magnetization behavior of the Fe-sheathed and bare wires. Obviously, these discrepancies are the consequences of the influence of the ferromagnetic sheath. We can divide these consequences into two categories: temperature dependent and temperature independent. It is natural to attribute the temperature independent consequences to the non-superconducting property of the iron sheath, whereas the temperature dependent consequences can be related to an interaction between the ferromagnetic sheath and the superconducting core.

To the first category, we can undoubtedly attribute only one temperature independent effect, which becomes obvious by direct comparison of the corresponding magnetization curves for the Fe-sheathed and bare wires. The difference is clearly seen below $B_s \simeq 0.34$~T marked by the dotted line in Fig.~\ref{prp}. $B_s$ is the saturation field of the magnetization for the ferromagnetic response of the Fe-sheath defined above $T_c$ in Fig.~\ref{iron} in a similar way as in Ref.~\cite{shield}. The magnetic behavior of the sheathed wire below $|B_s|$ exhibits zero magnetization for the ZFC branch and strongly suppressed (down to zero) magnetization for the FC branch. However, the behavior of the bare wire shows the largest values of the magnetization at $B_a \le |B_s|$ for the FC loops (masked by the flux-jumps below 15~K) and rapidly increasing diamagnetic magnetization for the ZFC branch. In the sheathed wire, up to $|B_s|$ all the magnetic energy within the volume of the wire is spent on alignment of magnetic domains inside the iron sheath and no energy is left to affect the superconductivity in the core. In other words, due to a magnetic permeability of the sheath a magnetic shielding effect occurs. The shielding in similar samples has been visualized in Ref.~\cite{shield} with the help of magneto-optical imaging. Note that in the ZFC curve of the shielded wire the Meissner state is shifted to a higher field, compared to the corresponding bare wire curve, by approximately 0.34~T. Therefore, the experimental first pseudo critical field $B_{c1}^{\rm ps}$ for the shielded wire can be defined as
\begin{equation}
B_{c1}^{\rm ps} \simeq B_{c1} + B_s \, .
\label{bc1} \end{equation}
The shift enables the first critical field $B_{c1} \sim 0.06$~T in MgB$_2$ superconductor to be exceeded by a factor of $\sim 5.6$.

To the second category of the $T$-dependent effects we can attribute the following features observed in Fig.~\ref{prp}: \\
(1) A clear overcritical state is observed at $T \ge 15$~K. This phenomenon is the most striking among all the observed effects. By the overcritical state here we imply that the maximum width of the hysteresis loops of the sheathed wire is always larger than the corresponding width of the magnetization in the bare wire. The width of the loops is well known to be directly proportional to the critical current density $J_c$.\\
(2) A suppression of the superconductivity occurs at $B_a > B_{\rm v}(T)$. $B_{\rm v}(T)$ is the field at which the overcritical state vanishes, i.e., the ascending branch of the sheathed wire crosses the corresponding branch of the bare wire (see Fig.~\ref{prp} for $B_{\rm v}$ definition). \\
(3) The irreversibility field of the sheathed wire is smaller than that of the bare wire at all temperatures (see Fig.~\ref{b}). \\
(4) The zero-field cooled (virgin) magnetization curve of the Fe-sheathed wire in increasing magnetic field is only shifted due to the iron screening effect, but its magnetization value in the minimum is comparable to the corresponding value of the bare wire, unlike the FC ascending branch case described in (1). \\ 
(5) The FC branches of the sheathed wire experience a strong magnetic history dependence. In contrast to the magnetization behavior of the bare wire, the ascending branch has much higher absolute values of the magnetization compared to those of the descending branch at all temperatures. \\
(6) The thermo-magnetic flux-jump instabilities in the sheathed wire appear over a wider field range and at higher temperatures than in the bare wire.

\subsection{Parallel orientation}\label{sub2}

The curves shown in Fig.~\ref{prl} exhibit magnetization behavior measured with the magnetic field applied parallel to the cylindrical axis of the wire at different temperatures. Apart from the orientation, the measurement conditions for the curves in each graph are identical with the corresponding conditions for the transverse orientation. However, one can clearly see significant discrepancies between the magnetization behaviors obtained for these two orientations (Figs.~\ref{prp} and \ref{prl}). 

The temperature independent effect of magnetic screening is still there due to the finite length of the sample, although it is far from being so prominent as in the transverse orientation. The saturation field of the ferromagnetic iron sheath $B_s \simeq 0.065$~T for the longitudinal orientation is much smaller than in the transverse case. Therefore, in agreement with Eq.~(\ref{bc1}), $B_{c1||}^{\rm ps} << B_{c1\perp}^{\rm ps}$. This magnetic anisotropy of $B_s$ is likely to be a consequence of the mechanical deformation process during the wire fabrication. Structurally, the process most probably elongates the iron grains in the sheath in the direction of drawing. Magnetically, the structural strains developed in the sheath during the mechanical deformation force the magnetization easy-axis of the iron to be parallel to the cylindrical axis of the wire.

\begin{figure}
\vspace{-0.5cm}
\includegraphics[scale=0.34]{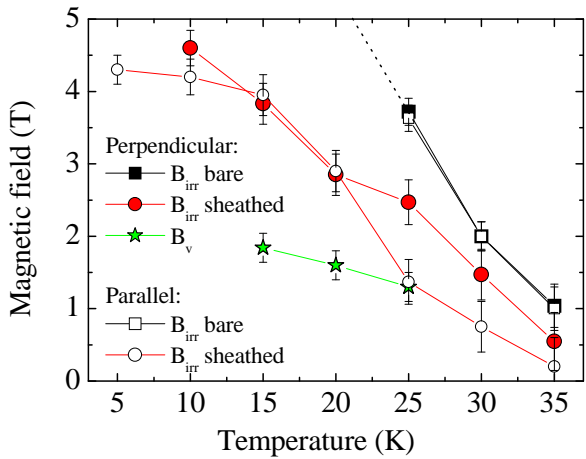}
\caption{\label{b}Irreversibility fields for the sheathed and bare wires in the parallel and perpendicular orientations. Due to the maximum field limitation of the SQUID magnetometer, the values of $B_{\rm irr}$ exceeding 5~T could not be reached. This is shown by the dotted line for the bare wire at $T < 25$~K.}
\end{figure}

As to the temperature-dependent effects, which arise due to the interaction between the magnetic sheath and the superconducting core, we point out the following peculiarities inherent to the field applied parallel to the wire (Fig.~\ref{prl}): \\
(1) No overcritical state is observed.\\
(2) The superconducting signal of the Fe-sheathed wire is suppressed compared to the signal of the bare wire over the entire field range. \\
(3) The irreversibility fields are smaller not only than in the bare wire, but also than in the case of the perpendicular orientation (Fig.~\ref{b}). \\
(4) The virgin curve behavior is similar to that described in case (4) of the transverse orientation.  \\
(5) The magnetic history dependent effect is similar to that described in (5) for the transverse case. \\
(6) The thermo-magnetic flux-jump instabilities are also similar to case (6) for the perpendicular orientation. 

\section{Discussion}

The temperature independent case of the iron screening effect is rather straightforward for both field orientations, and it has been described and visualized in our previous work \cite{shield}. In this work, we deal with the temperature-dependent phenomena which arise from the interaction between the sheath and the superconducting core.

\subsection{Overcritical state}\label{discoc}

\begin{figure}
\includegraphics[scale=0.45]{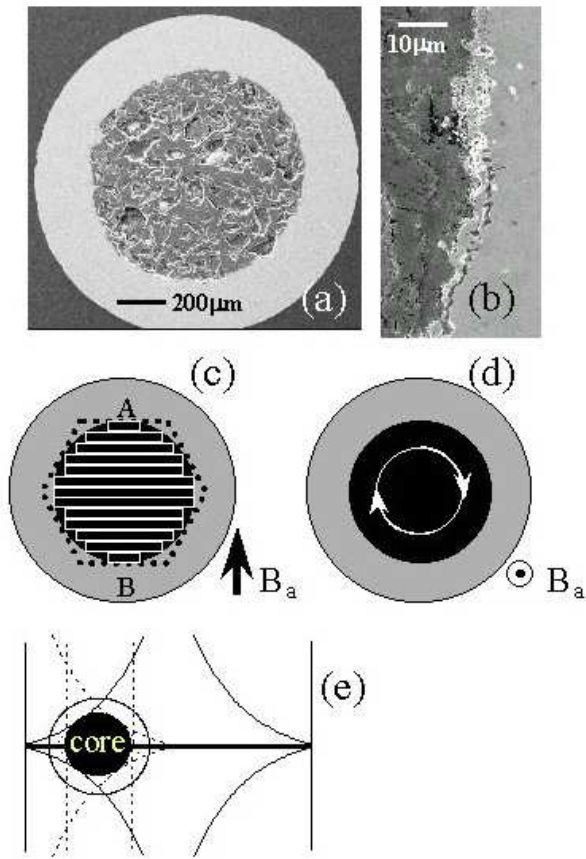}
\caption{\label{sem}(a)A cross section of an MgB$_2$ wire sheathed in iron obtained with the help of scanning electron microscopy. The diameter of the sheathed wire is $\simeq 1.5$~mm, while the diameter of the core is $\simeq 1.0$~mm. (b) An interface area at high magnification exhibiting a 6.2-9.1~$\mu$m thick white layer of MgO. The electro-magnetic schematics of the sheathed wire cross section are shown for transverse (c) and longitudinal (d) field orientations. The superconducting core in the transverse orientation can be approximated as a stack of strips with their own Meissner shielding currents confined within each strip (c). The dotted line in (c) shows a hexagonal approximation of the core-sheath interface. The white circular arrows in (d) denote shielding currents for the longitudinal orientation. (e) Similarities of different magnetic environments for the wire and an isolated strip (see text).}
\end{figure}

The most scientifically interesting and practically important effect obtained in this work is the overcritical state observed in the case of the applied field perpendicular to the sheathed wire axis (Fig.~\ref{prp}). In general, the existence of the overcritical state as a result of an interaction between a superconducting {\it strip} and a soft magnetic environment was predicted in Ref.~\onlinecite{gen}. The main idea is that for some environment configurations the super-current profile over the strip can be redistributed so that the super-current is pushed from the flux-filled strip edges to the central flux-free part of the strip, resulting in a total overcritical current $I^{oc} > I_c$. The most pronounced enhancement of the current was found for an open convex magnetic cavity. The cylindrical cavity was shown to be a less favorable configuration for a current enhancing redistribution. Note that in our case the core is situated inside a cylindrical cavity (see Fig.~\ref{sem}(a)). It was further shown \cite{gen1} that if a strip is squeezed between two parallel magnets, forming a ``sandwich"-like configuration, it can lead to a significant {\em undesirable} current enhancement at the strip edges. The enhancement at the edges would eventually lead to an overall $I_c$ degradation in the strip, which eliminates any possibility for the occurrence of an overcritical state. In our case, the sheath can also be considered to be parallel to the core in the vicinity of points A and B (see Fig.~\ref{sem}(c)). The super-current redistribution is possible in the strips since the Meissner shielding currents can flow over the entire surface of the thin superconductor \cite{gen}. The only controversial experimental indication for the occurrence of the overcritical state occurrence in strips has been recently reported for YBa$_2$Cu$_3$O$_7$ films by employing the magneto-optical imaging technique \cite{jar}. Overcritical Meissner currents were locally obtained by inversion of the Biot-Savart law in the region next to magnets set perpendicular to the film but not on the middle of the film which contradicts to the theoretical prediction \cite{gen}. Moreover, no global investigations, such as magnetization or transport measurements, have been carried out. Indeed, the geometrical configuration for which the overcritical state was predicted to exist in strips is extremely difficult to realize in practice.

The situation is completely different in our case of the bulk-like wire, which is a realistic, easy to manufacture configuration of MgB$_2$ round wires sheathed by iron. In the superconducting core of the wire the super-currents flow only within the magnetic field penetration depth $\lambda$, unless the superconductor is filled with the magnetic flux. According to Ref.~\onlinecite{gen} the flux-free Meissner state is physically important condition for overcritical currents to exist. Nevertheless, in spite of being bulk-like, having direct contact with parallel magnets and carrying out our experiments on the wires in FC state we are able to clearly show that the overcritical state does exist in the wires (Fig.~\ref{prp}).

\begin{figure}
\vspace{-0.5cm}
\includegraphics[scale=0.43]{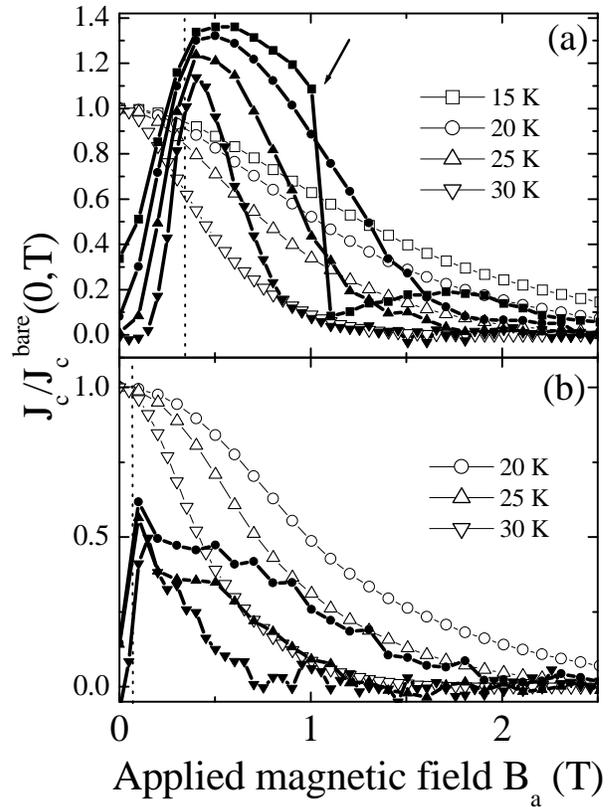}
\caption{\label{jc}Normalized critical current densities calculated from measured magnetization loops for (a) perpendicular and (b) parallel field orientation. The large open symbols are used for the curves measured for the bare wire, whereas the smaller solid symbols and thick connecting lines denote the behavior of the sheathed wire. The arrow points at a flux-jump. The dotted lines mark $B_s$ for both orientations.}
\end{figure}

In figure~\ref{jc}(a) we directly compare critical current densities of the bare and sheathed wires, calculated from the critical state model for a cylinder in a perpendicular magnetic field. As can be clearly seen, in a certain magnetic field interval varying with temperature the (over)critical current density $J_c^{oc}$ of the sheathed wire is significantly higher than $J_c$ for the bare wire. Moreover, its maximum value $J_{c, {\rm max}}^{oc}$, occurring at the field $B_a^{oc}$ which is slightly larger than $B_{s}$, is even higher than $J_c(0,T)$. Unfortunately, it is impossible to compare critical current densities at $B_a < |B_s|$, since due to the magnetic screening effect a rapidly vanishing superconducting signal is measured. As a result no real $J_c^{oc}$ at $B_a < |B_s|$ is available in the case of magnetic measurements.

To explain the overcritical state developing in a round wire confined in a cylindrical magnetic environment (Fe-sheath) we have constructed a phenomenological model which can be described with the help of the schematic in Fig.~\ref{sem}(c). In the case of the field applied perpendicular to the wire axis we assume that Meissner shielding currents flow within quasi two-dimensional planes. The wire can then be considered as a stack of strips tightly packed in the cylindrical cavity. Indeed, the relationship between magnetic field and current density in an isolated ``bulk" strip in a perpendicular $B_a$ was shown to be easily transformed to the corresponding relation in the strip-array system \cite{mawa} with each strip having a thickness equal to the effective penetration depth for these strips \cite{pearl}. The possibility of this transformation also implies that the field-current behavior of the entire ``bulk" core can be derived from that of an isolated, centrally situated strip.

In our case, the strips at the top (A) and bottom (B) of the cavity (Fig.~\ref{sem}(c)) experience the undesirable parallel magnet configuration with only one side exposed to the Fe-sheath (magnet). On the other hand, these strips shield the centrally situated ones from the parallel magnet influence. The central strips would then experience only the interaction with the Fe-sheath located at the sides of the strips, which would form the geometry of an open concave cavity with the opposite curvature to the convex configuration. However, one can consider the unified linear approximation shown in Fig.~\ref{sem}(c,e) for the curvatures of the concave and convex cavities. More similarities between the wire and the strip for different magnetic environment configurations are shown in Fig.~\ref{sem}(e). The diameter of the wire and the width of the strip are kept at the 1:5 ratio to preserve the ratio between the wire and the strip considered in this work and in Ref.~\cite{gen}, respectively. Thus, it is legitimate to expect to obtain similar values of the current enhancement in both cases. The enhancement of the critical current density to the maximum of the overcritical current density $J_c^{oc}$ obtained for the case of the convex cavity is given by \cite{gen}
\begin{equation}
\frac{J_c^{oc}}{J_c} = \left[ \frac{1}{4\pi} \tan \left( \frac{\pi D}{4a} \right) \right]^{1/2}\, .
\label{max} \end{equation}
$D \simeq 1.0$~mm is the superconducting core diameter and $a \simeq 6.2-9.1$~$\mu$m is the distance between the core and the sheath. As revealed by scanning electron microscopy (SEM), a layer of presumably MgO is formed on the interface between the Fe-sheath and the MgB$_2$ core (Fig.~\ref{sem}(b)) \cite{PhysC2}. Substituting the maximum $J_{c, {\rm max}}^{oc}/J_c = 1.6$ enhancement ratio obtained in the experiment at $B_a^{oc}$ and $T < 30$~K, we can find a ratio $D/a$ and compare it to the one obtained with the help of SEM. It turned out that $D/a \sim 112$ obtained from substituting $J_{c, {\rm max}}^{oc}/J_c = 1.6$ into Eq.~(\ref{max}) is in a good agreement with the dimensional ratio obtained by SEM investigations. This fact can indicate the similar origins of the current enhancement in our case and in the case of a superconducting strip \cite{gen}.

\begin{figure}
\vspace{-0.5cm}
\includegraphics[scale=0.34]{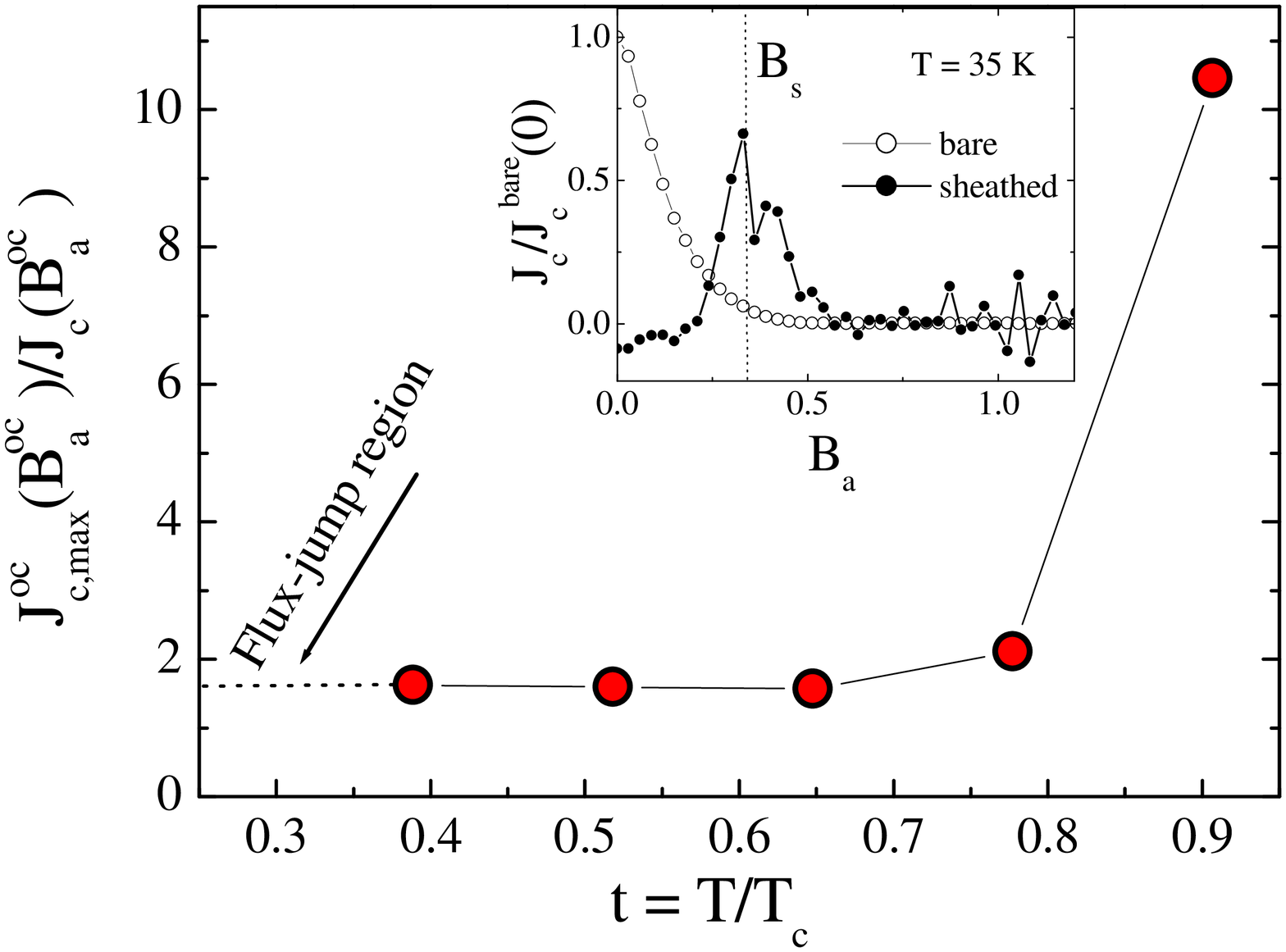}
\caption{\label{oc}Critical current density enhancement at $B_a = B_a^{oc}$ at which the maximum enhancement occurs as a function of the normalized temperature. The dotted line marks the expected enhancement at lower temperatures overshadowed by flux-jump instabilities. The inset shows the normalized $J_c(B_a)$ graph for $T = 35$~K in the perpendicular orientation, similar to those exhibited in Fig.~\ref{jc}(a).}
\end{figure}

It is worth noting that according to Eq.~(\ref{max}) the $J_{c, {\rm max}}^{oc}/J_c$-ratio is independent of temperature. This independence is also observed in our experiment at temperatures below 30~K (Fig.~\ref{oc}). The behavior of the $J_c$ enhancement below $T = 15$~K which is overshadowed by the flux-jump instabilities is expected to remain constant as shown by the dotted line in Fig.~\ref{oc}. Strikingly, the ratio is drastically boosted up to one order of magnitude at temperatures approaching $T_c$. This effect we attribute to the consequences of the magnetic screening effect. Indeed, at $T \ge 30$~K, $B_{\rm irr}$ in the sheathed wire is rapidly approaching $B_s$. Below $B_s$, the core is unaffected by field alteration. Within $B_s < B_a < B_{\rm irr}$, the wire experiences the interplay between the rapid $J_c^{oc}$ enhancement and the proximity of $B_{\rm irr}$, leading to a sharp $J_c^{oc}$ peak near $B_{\rm irr}$ (see the inset in Fig.~\ref{oc}). On the contrary, the $J_c$-dependence of the bare wire monotonically decreases and vanishes at $B_{\rm irr}$. These differences in the $J_c(B_a)$ behavior lead to large enhancement values in the vicinity of $T_c$.

In the case of the field applied parallel to the wire axis (Fig.~\ref{sem}(d)), no overcritical state is measured (Figs.~\ref{prl} and \ref{jc}). This is because for this orientation no super-current redistribution takes place due to cylindrical symmetry (Fig.~\ref{sem}(d)). In the fields $B_a > B_{c1||}^{\rm ps}$ the Fe-sheathed samples experience the well-known Bean-like field and current profiles similar to those in the bare wire. This is also in agreement with the prediction in Ref.~\onlinecite{gen}. In fact, $J_c$ of the bare wire is higher than that of the sheathed wire (Fig.~\ref{jc}(b)), which might be associated with the local suppression of the superconductivity by the magnetic material. The difference between perpendicular and parallel field orientation is a strong support in favor of the super-current redistribution from edges (of the strips depicted in Fig.~\ref{sem}(c)) towards the central area of the superconducting core.

A further support of such the redistribution has recently been provided by magneto-optical (MO) imaging \cite{mo}. The modified flux profile induced by the transport super-current interaction with the externally magnetized sheath has been visualized within the MgB$_2$ core of the Fe-sheathed wire in a similar way as it was done in Ref.~\onlinecite{shield}. The observed flux profile within the core was compatible with the unusual enhancement of the super-current in the middle of the core \cite{mo}. This effect was observed only in the case of the wire cooled in an externally applied field (FC state). In addition, it is worth noting that recent theoretical results have shown that in a bulk round superconductor wire placed in a semi-infinite cylindrical magnet the sheet currents in the Meissner state can be enhanced as a result of redistribution from the edges towards the center of the superconductor \cite{yam}.

A magnetic history-dependent effect with a strongly pronounced asymmetry between the FC ascending and descending branches of the magnetization is observed in the sheathed wires for both field orientations (Figs.~\ref{prp} and \ref{prl}). This asymmetry arises due to the interplay between the irreversibility induced by pinning and the sheath permeability. Indeed, the ascending branch always exhibits much higher values of the magnetization compared to the descending branch. For example, at $T = 15$~K the maximum absolute value of the ascending branch at $B_a \simeq 0.5$~T is higher than the corresponding value of the descending branch by a factor of $\simeq 5$ for the transverse case (Fig.~\ref{prp}). This factor decreases with increasing temperature, whereas the maximum value of the descending branch remains almost constant. At lower temperatures, one can anticipate the maximum value of the ascending branch to be even higher and the difference between the branches even larger, but the magnetization behavior is strongly overshadowed by the thermo-magnetic flux-jump instabilities. Thus, the ascending branch behavior is attributable to the superconducting properties, whereas the behavior of the descending branch is basically controlled by the magnetized sheath. Note that the signal of the descending branch is comparable for both field orientations.

The influence of the sheath magnetization also becomes dominating for the ascending branches above $B_{\rm v}(T)$. Eventually, the suppression of the superconductivity by the magnetized sheath leads to significantly smaller values of the irreversibility fields compared to the bare wires (Fig.~\ref{b}).

\subsection{Virgin ZFC magnetization curve behavior}\label{discv}

As mentioned above and clearly seen in Figs.~\ref{prp} and \ref{prl}, the virgin ZFC magnetization curves for the sheathed wire are shifted by $\simeq B_s$ with respect to the corresponding curves for the bare wire over the entire temperature range. The interval of zero-magnetization signal arising from the magnetic screening effect \cite{dj} is referred to as the extended \cite{shield} or pseudo-Meissner effect \cite{sumpt}. However, we would like to emphasize that in the case of the transverse field the absolute value of the magnetization at its minimum reproduces the corresponding value obtained for the bare wire (Fig.~\ref{prp}, see $T \ge 20$~K). No magnetization enhancement is observed, in contrast to the FC ascending branch discussed in the previous subsection. This implies that as long as no frozen flux is present in the superconducting core, no enhancement takes place. This is because Miessner shielding currents are induced in the sheathed core only within the field penetration depth at $B_a \ge B_s$. In other words, if no flux is frozen in the core, no super-currents are present inside the bulk of the core in agreement with the Maxwell equations. This means that no super-current redistribution is possible. Therefore, the model proposed in the previous subsection for the explanation of the overcritical state explanation is valid only for a superconductor {\it filled} with magnetic flux. This conclusion is also in agreement with the results of the recent MO experiments \cite{mo} mentioned in the previous subsection. The flux density in the center of the core should be rather diluted just above $|B_s|$ upon increasing the field from $B_a = 0$~T in the FC state. This diluted flux can enable the super-current redistribution as predicted in Refs.~\cite{gen,yam} for the Meissner state. However, a further investigation has to be done in order to understand how the presence of the flux in the superconductor comply with the theory.

In the case of the longitudinal field (Fig.~\ref{prl}) the value of the magnetization minimum in the virgin curves of the sheathed wire is significantly suppressed. This can also be explained by suppression of the superconductivity by the magnetic sheath. This would agree with the general explanation for the magnetization behavior at this orientation discussed in the previous subsection.

\subsection{Flux-jump instabilities}\label{discfj}

As we have already mentioned previously \cite{sust4}, for MgB$_2$ wires prepared by the {\it in-situ} method the presence of the iron sheath leads to strong thermo-magnetic flux-jump instabilities \cite{inst}. Indeed, as can be seen in Fig.~\ref{prp}, the flux-jumps appear for the Fe-sheathed wire at $T \le 15$~K, whereas for the bare wire the flux-jumps are absent at $T = 15$~K. Moreover, the flux-jumps in the Fe-sheathed wire extend over a much broader field range than in the bare wire at $T \le 10$~K. In the case of the field applied parallel to the wire axis, the differences between flux-jump behaviors of the sheathed and bare wires are present, as well (Fig.~\ref{prl} for $T \le 15$~K). Generally, the low thermal conductivity of the iron sheath worsens the thermal sink environment of the superconductor, compared to the environment of the steady cold helium gas flow around the bare core. As a consequence, the sheathed system is more sensitive to local overheating due to vortex avalanches \cite{inst}. Sheath materials, such as copper and silver, which could stabilize the wires \cite{stab}, have been shown to be too soft to support an acceptable level of critical currents and were chemically incompatible with the {\it in-situ} MgB$_2$ core preparation \cite{pan}.

Furthermore, taking into account the fact that the overcritical state produces $J_c^{oc}$ which is significantly higher than $J_c$ in the bare wires, one can presume that a stronger pronounced flux-jump behavior in the sheathed wires might also be a result of an interplay between a worsened heat sink environment and the higher $J_c^{oc}$ induced by the sheath. Generally, the higher $J_c$, the wider is the field range for the instabilities \cite{inst}. Hence, one might expect a stronger pronounced flux-jump behavior over a wider field range for the wire in a perpendicular field with $J_c^{oc} > J_c$ of the wire in a parallel field. In fact, the flux-jump behavior is quite similar for both orientations (Figs.~\ref{prp} and \ref{prl}). This fact might indicate that the Meissner shielding super-currents near the core surface are similar for both orientations. Therefore, the condition for the flux-jump appearance, basically given by local violation of the Ampere's law
\begin{equation}
\frac{\partial B_a}{\partial x} > - \mu_0 J_c
\label{amp} \end{equation}
(where $\mu_0$ is the permeability of the free space and $x$ is the coordinate along the flux gradient), would be controlled by the shielding $J_c$ near the surface, but not by $J_c^{oc}$ which is achieved in the bulk of the flux-filled core as a result of the super-current redistribution in the perpendicular fields.
 
\section{Conclusion}

In summary, we showed that an overcritical current state exists in round superconducting MgB$_2$ wires sheathed in iron with $J_c^{oc} > J_c$ by one order of magnitude at temperatures close to $T_c$. At lower temperatures the enhancement is temperature independent with $J_{c, {\rm max}}^{oc}/J_c \simeq 1.6$. A phenomenological model valid for a round core filled with magnetic flux has been proposed which uncovers analogies between overcritical states in the round wires investigated in this work and in the strips treated theoretically in Refs.~\onlinecite{gen1,gen}. (i) The experimental temperature independence of $J_{c, {\rm max}}^{oc}/J_c$ below $t \simeq 0.8$, (ii) the overcritical current density estimation with the help of Eq.~(\ref{max}), (iii) the magnetic history dependence of the magnetization for ZFC and FC states and (iv) particularly absence of the overcritical state for the parallel field orientation are the main points which indicate that the origin of the overcritical state existence in the Fe-sheathed bulk wires is similar to that predicted for the thin strips. The overcritical state has been accounted for as a result of magnetic interactions between the permeable iron sheath and the superconducting core. These interactions in perpendicular fields can lead to the redistribution of super-currents within the flux-filled superconducting core, resulting in the appearance of the overcritical state. We anticipate that the effects observed should be present not only for the Fe-MgB$_2$ sheath-core wire system investigated, but also for any kind of compatible wire composition comprising a superconducting round core and a soft magnetic sheath material.

The interaction of the iron sheath and superconductor has also been shown to lead to suppression of the superconductivity, in particular to degradation of the irreversibility field, in the cases when the sheath is fully magnetized.

The effect of flux-jump instabilities is shown to be significantly enhanced by the Fe-sheath. Similar flux-jump behaviors in the sheathed wires for both orientations can suggest that $J_c$ near the surface of the core is of a similar value for both cases. This would again indicate that the overcritical currents, in the case of a transverse applied field, are achieved in the bulk of the superconducting core.

\begin{acknowledgments}
We would like to thank T. Silver for careful reading of the manuscript and valuable critical remarks, as well as E. W. Collings, Hyper Tech Research Inc, and Alphatech Ltd for support. This work is financially supported by the Australian Research Council.
\end{acknowledgments}


\end{document}